\documentclass[aps,showpacs,twocolumn,amsmath,amssymb,superscriptaddress,prl]{revtex4-1}

\usepackage{graphicx}
\usepackage{color}

\usepackage{bm}

\usepackage{comment}

\begin{document}

\title{Violation of local reciprocity in charge--orbital interconversion}

\author{Hisanobu Kashiki}
\affiliation{Department of Applied Physics and Physico-Informatics, Keio University, Yokohama 223-8522, Japan}

\author{Hiroki Hayashi}
\affiliation{Department of Applied Physics and Physico-Informatics, Keio University, Yokohama 223-8522, Japan}

\author{Dongwook Go}
\affiliation{Peter Gr\"unberg Institut and Institute for Advanced Simulation, Forschungszentrum J\"ulich and JARA, 52425 J\"ulich, Germany}
\affiliation{Institute of Physics, Johannes Gutenberg University Mainz, 55099 Mainz, Germany}
\affiliation{Department of Physics, Korea University, Seoul 02841, Republic of Korea}

\author{Yuriy Mokrousov}
\affiliation{Peter Gr\"unberg Institut and Institute for Advanced Simulation, Forschungszentrum J\"ulich and JARA, 52425 J\"ulich, Germany}
\affiliation{Institute of Physics, Johannes Gutenberg University Mainz, 55099 Mainz, Germany}

\author{Kazuya Ando
\footnote{Correspondence and requests for materials should be addressed to ando@appi.keio.ac.jp}
}
\affiliation{Department of Applied Physics and Physico-Informatics, Keio University, Yokohama 223-8522, Japan}
\affiliation{Keio Institute of Pure and Applied Sciences, Keio University, Yokohama 223-8522, Japan}
\affiliation{Center for Spintronics Research Network, Keio University, Yokohama 223-8522, Japan}

\begin{abstract}
We demonstrate a violation of local reciprocity in the interconversion between charge and orbital currents. By investigating orbital torque and orbital pumping in W/Ni bilayers, we show that the charge-orbital interconversion in the bulk of the W layer exhibits opposite signs in the direct and inverse processes---the direct and inverse orbital Hall effects being positive and negative, respectively. This finding provides direct evidence of local nonreciprocity in the charge-orbital interconversion, in agreement with a theoretical prediction. These results highlight the unique characteristics of charge-orbital coupled transport and offer fundamental insights into the mechanisms underlying orbital-current-driven phenomena.
\end{abstract}

\maketitle

Since the discovery of the spin Hall effect (SHE), the interconversion between charge and spin currents through spin-orbit coupling has played a crucial role in the development of spintronics~\cite{RevModPhys.87.1213}. The direct process of the SHE converts a charge current into a spin current, enabling the electric control of magnetization through spin-orbit torques~\cite{AndoPRL,liu2012spin}. 
This process has attracted extensive attention for its topological and quantum mechanical nature, as well as its potential for spintronic applications~\cite{RevModPhys.91.035004}.
The inverse process of the SHE, the conversion from a spin current into a charge current, offers a way for the electric detection of spin currents~\cite{saitoh2006conversion,valenzuela2006direct,KimuraPRL}. This mechanism has led to the discovery of various spintronic phenomena, including spin pumping---a process in which magnetization dynamics generates a spin current~\cite{RevModPhys.87.1213}.

\begin{figure}[tb]
\includegraphics[scale=1]{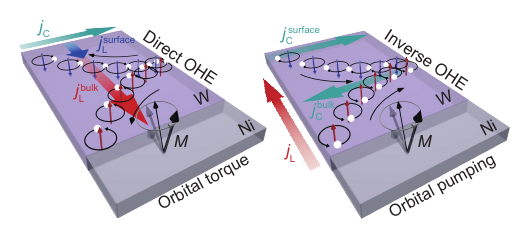}
\caption{Schematic illustration of orbital torque induced by the direct OHE (left) and the inverse OHE induced by orbital pumping (right) in W/Ni. In the direct OHE, $j_\mathrm{C}$ represents the applied charge current, while $j_\mathrm{L}^\mathrm{bulk}$ and $j_\mathrm{L}^\mathrm{surface}$ denote the orbital currents generated by the bulk and surface contributions, respectively. The red and blue colors represent the positive and negative orbital polarizations, respectively. In the direct OHE, $j_\mathrm{L}^\mathrm{bulk}$ and $j_\mathrm{L}^\mathrm{surface}$ carry opposite orbital angular momentum because the bulk and surface orbital Hall conductivities have opposite signs in W. 
In the inverse OHE, $j_\mathrm{L}$ is the injected orbital current, while $j_\mathrm{C}^\mathrm{bulk}$ and $j_\mathrm{C}^\mathrm{surface}$ are the charge currents generated by the bulk and surface contributions, respectively. These charge currents, $j_\mathrm{C}^\mathrm{bulk}$ and $j_\mathrm{C}^\mathrm{surface}$, flow in opposite directions. $M$ denotes the magnetization.
}
\label{fig_schematic} 
\end{figure}

While spintronics has primarily focused on the generation and conversion of spin currents, recent advances have revealed that charge currents can also drive the nonequilibrium dynamics of orbital angular momentum (OAM). A key phenomenon underlying OAM dynamics is the direct orbital Hall effect (OHE)---the orbital counterpart of the direct SHE---in which a charge current generates an orbital current, i.e., a flow of electrons carrying nonzero OAM (see Fig.~\ref{fig_schematic})~\cite{PhysRevLett.95.066601,PhysRevLett.102.016601,PhysRevLett.121.086602,PhysRevB.102.035409,PhysRevLett.126.056601}. Theoretical and experimental progress on the direct and inverse OHEs has led to the emergence of orbitronics, which aims to utilize the interplay between charge and orbital currents in solid-state devices~\cite{go2021orbitronics,KIM2022169974,2025Andoreview}.

One of the fundamental challenges in harnessing charge-orbital coupled transport is to understand the reciprocity of charge-orbital interconversion~\cite{PhysRevLett.132.226704,4c06056,gao2025nonlocalelectricaldetectionreciprocal}. 
In the OHE, the local, layer-resolved charge-to-orbital and orbital-to-charge conversion processes are characterized by the local direct and inverse orbital Hall conductivities, $ \sigma_{\mathrm{OH}}^{\mathrm{dir}}(z)$ and $ \sigma_{\mathrm{OH}}^{\mathrm{inv}}(z)$, at position $z$, with the orbital current flowing along the $z$ direction~\cite{supplementary}. 
The global response, obtained by integrating over the film thickness, is governed by Onsager's reciprocal relations, which require 
\begin{equation}
\int \sigma_{\mathrm{OH}}^{\mathrm{dir}}(z) dz = \int  \sigma_{\mathrm{OH}}^{\mathrm{inv}}(z) dz,\label{global}
\end{equation}
a relation that has been rigorously proven for the proper orbital current~\cite{go2024local}. 
On the other hand, local reciprocity, which concerns whether the equality $\sigma_{\mathrm{OH}}^{\mathrm{dir}}(z) = \sigma_{\mathrm{OH}}^{\mathrm{inv}}(z)$ holds at each position $z$, is nontrivial. 
This is because Onsager's reciprocal relations are derived as a corollary of the fluctuation-dissipation theorem, which relates macroscopic responses in nonequilibrium to microscopic correlations in equilibrium. While this guarantees reciprocity for global, spatially integrated responses, it does not impose constraints on local responses. 
In fact, first-principles calculations for W have explicitly demonstrated a violation of local reciprocity between direct and inverse processes,
\begin{equation} 
\sigma_{\mathrm{OH}}^{\mathrm{dir}}(z) \neq \sigma_{\mathrm{OH}}^{\mathrm{inv}}(z),\label{local} 
\end{equation}
even though the global reciprocity condition in Eq.~(\ref{global}) is satisfied~\cite{go2024local}. 
This local nonreciprocity originates from the non-conservation of OAM due to its strong coupling with the lattice through crystal fields. Such coupling is accounted for by defining the orbital current as the sum of the conventional orbital current and torque dipole, the latter being associated with the interaction between OAM and the crystal lattice~\cite{go2024local}. The torque dipole is particularly significant at interfaces, where the anisotropic crystal potential efficiently mediates angular momentum exchange between the electrons and the lattice, strongly influencing the local profile of charge-orbital interconversion in the W layer. As a consequence, local charge currents induced by orbital currents deviate from the expectation of strict reciprocity. Notably, in $\alpha$-W, the local responses exhibit opposite signs between the direct and inverse OHEs, as summarized in Table~I in Appendix~A.
However, despite recent observations of the direct and inverse OHEs~\cite{Cr-orbital,Ta-orbital,PhysRevResearch.4.033037,hayashi2022observation,choi2021observation,hayashi2022observation,hayashi2023orbital,PhysRevB.107.134423,xu2024orbitronics,hayashi2024observation,seifert2023time,gao2024control,4c06056}, the nature of their reciprocal relationship has remained unclear.

In this Letter, we report the observation of a violation of local reciprocity in the charge-orbital interconversion in $\alpha$-W. 
Because of the predicted opposite signs between the bulk direct and inverse orbital Hall conductivities~\cite{go2024local}, W is a suitable platform for testing local nonreciprocity~\cite{supplementary}.
In nanoscale spintronic and orbitronic devices, where measured quantities can deviate from macroscopic averages, local reciprocity can be investigated by systematically varying the device thickness. This approach provides a route to exploring coupled transport phenomena beyond the framework of Onsager's reciprocity. Through systematic measurements of orbital torque and orbital pumping in W/Ni bilayers, we demonstrate that the OHE in the W bulk exhibits opposite signs in the direct and inverse processes, revealing the unique nature of charge-orbital coupled transport.

\begin{figure}[tb]
\includegraphics[scale=1]{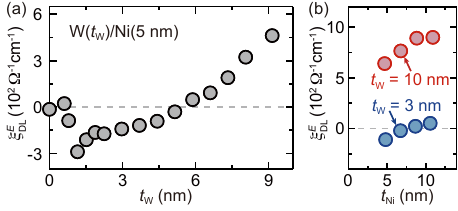}
\caption{The damping-like torque efficiency $\xi_\mathrm{DL}^E$ determined by ST-FMR for the W/Ni bilayers. (a) Dependence of $\xi_\mathrm{DL}^E$ on the W thickness $t_\mathrm{W}$ in the W($t_\mathrm{W}$)/Ni(5~nm) bilayer. The values of $t_\mathrm{W}$, determined from the sputtering rate, are as follows: 0, 0.61, and 0.79~nm; in 0.36~nm steps from 1.15 to 2.24~nm; in 0.73~nm steps from 2.97 to 8.06~nm; and 9.15~nm. (b) Dependence of $\xi_\mathrm{DL}^E$ on the Ni thickness $t_{\rm Ni}$ in the W(3~nm)/Ni($t_\mathrm{Ni}$) (blue) and W(10~nm)/Ni($t_\mathrm{Ni}$) (red) bilayers. The values of $t_\mathrm{Ni}$, determined from the sputtering rate, range from 4.81 to 10.58~nm in steps of 1.92~nm.}
\label{fig_STFMR} 
\end{figure}

We measured the current-induced torque in W/Ni bilayers using spin-torque ferromagnetic resonance (ST-FMR) to investigate the charge-to-orbital conversion in W. The samples were fabricated by magnetron sputtering with the structure of SiO$_2$(4~nm)/W($t_\mathrm{W}$)/Ni($t_\mathrm{Ni}$)/SiO$_2$-substrate, where the numbers in parentheses indicate the layer thicknesses. The measurement and analysis procedures follow those in our previous work~\cite{supplementary,hayashi2022observation}. 
Figure~\ref{fig_STFMR}(a) shows the W-layer thickness $t_\mathrm{W}$ dependence of the damping-like torque efficiency, 
\begin{equation}
\xi_\text{DL}^E = \left(\frac{2e}{\hbar }\right)\frac{\mu_{0}M_\text{s}t_\text{Ni}H_\text{DL}}{E}, \label{eq:DLeff}
\end{equation}
determined by ST-FMR for the W($t_\mathrm{W}$)/Ni(5~nm) bilayers, where $e$ is the elementary charge, $\hbar$ is the reduced Planck constant, $M_\mathrm{s}$ is the saturation magnetization, $H_\mathrm{DL}$ is the damping-like effective field, and $E$ is the applied electric field. 
At $t_\mathrm{W} < 5$~nm, $\xi_\text{DL}^E$ with the negative sign appears as $t_\mathrm{W}$ increases. This behavior is consistent with the torque arising from the direct SHE in the W layer. In a simple spin diffusion model, the SHE-induced $\xi_\text{DL}^E$ scales as $1-\mathrm{sech}(t_\mathrm{W}/\lambda_\mathrm{S})$, where $\lambda_\mathrm{S}$ is the spin diffusion length~\cite{PhysRevLett.106.036601}. This model predicts that SHE-induced $\xi_\text{DL}^E$ increases with $t_\mathrm{W}$ and saturates when $t_\mathrm{W}$ exceeds approximately $3\lambda_\mathrm{S}$. The negative sign of $\xi_\text{DL}^E$ aligns with that of the direct SHE in the W layer~\cite{PhysRevMaterials.6.095001}, and the rapid increase in $\xi_\text{DL}^E$ with $t_\mathrm{W}$ is consistent with the short spin diffusion length of W~\cite{PhysRevMaterials.2.014403}, suggesting that the direct SHE is the primary mechanism driving the torque at $t_\mathrm{W} < 5$~nm.

Figure~\ref{fig_STFMR}(a) shows that $\xi_\text{DL}^E$ changes sign from negative to positive as $t_\mathrm{W}$ increases, indicating a crossover in the dominant mechanism responsible for the observed torque.
Complementary second-harmonic Hall measurements confirm the reliability of $\xi_\mathrm{DL}^E$ determined by ST-FMR, including the sign reversal with increasing $t_\mathrm{W}$~\cite{hayashi2022observation}. 
At $t_\mathrm{W} > 5$~nm, $\xi_\text{DL}^E$ increases with $t_\mathrm{W}$, suggesting that an additional torque with the positive sign becomes dominant at the larger $t_\mathrm{W}$. The observed $t_\mathrm{W}$-dependent behavior of this additional torque suggests that it originates from the charge current flowing in the bulk of the W layer, rather than from the SiO$_2$/W, W/Ni, and Ni/SiO$_2$ interfaces~\cite{supplementary}.
 We note that the positive sign of the additional torque is consistent with the sign of the direct OHE in W~\cite{PhysRevMaterials.6.095001}, suggesting that it can be attributed to the bulk direct OHE. In analogy to the model of the SHE-induced torque, the orbital torque due to the direct OHE in the W layer may also scale as $1-\mathrm{sech}(t_\mathrm{W}/\lambda_\mathrm{L})$, where $\lambda_\mathrm{L}$ is the characteristic orbital transport length. The observed $t_\mathrm{W}$-dependence of the additional torque suggests that $\lambda_\mathrm{L}$ is larger than $\lambda_\mathrm{S}$, consistent with previous reports~\cite{hayashi2022observation,PhysRevResearch.4.033037,choi2021observation}. In this case, the orbital Hall contribution to $\xi_\text{DL}^E$ from the bulk charge current is negligible at small $t_\mathrm{W}$. However, this contribution increases with $t_\mathrm{W}$ and can exceed that from the SHE at sufficiently large $t_\mathrm{W}$, resulting in the sign reversal of $\xi_\text{DL}^E$ with $t_\mathrm{W}$.

A key difference between spin and orbital phenomena lies in the characteristic length scales associated with the thickness of the ferromagnet (FM)~\cite{PhysRevLett.130.246701}. 
When a spin current is injected into a FM, it decays rapidly---typically within 1~nm---due to spin dephasing. Because of this short-range nature of spin transport, the damping-like effective field due to spin injection scales as $H_\mathrm{DL} \propto 1/t_\mathrm{Ni}$. Consequently, the damping-like torque efficiency defined in Eq.~(\ref{eq:DLeff}) is independent of the FM-layer thickness once it exceeds the spin dephasing length, as demonstrated in spin-Hall-dominated Pt/Ni$_{81}$Fe$_{19}$ bilayers~\cite{supplementary}.
In contrast, orbital currents can propagate over much longer distances than spin dephasing length through momentum-space hotspots that suppress oscillations of OAM~\cite{PhysRevLett.130.246701}. Owing to this long-range nature of orbital transport in FMs, $\xi_\mathrm{DL}^E$ arising from orbital torque increases with the FM-layer thickness~\cite{PhysRevLett.130.246701,hayashi2022observation,PhysRevLett.128.067201,PhysRevB.107.134423,PhysRevResearch.4.033037,PhysRevB.105.104434,PhysRevResearch.5.023054,gao2024control}.

The significant orbital contribution to $\xi_\mathrm{DL}^E$ is supported by a long-range nature of $\xi_\mathrm{DL}^E$. 
As shown in Fig.~\ref{fig_STFMR}(b), $\xi_\mathrm{DL}^E$ increases with $t_\mathrm{Ni}$ in the W(10~nm)/Ni($t_\mathrm{Ni}$) bilayer, even though the $t_\mathrm{Ni}$ range is much larger than the spin dephasing length. In this $t_\mathrm{Ni}$ range, we also confirmed that $\xi_\mathrm{DL}^E$ is negligible in Ni single layers~\cite{supplementary}. These results indicate that $\xi_\mathrm{DL}^E$ with the positive sign is dominated by the orbital torque induced by the bulk direct OHE. 
Figure~\ref{fig_STFMR}(b) also shows that in the W(3~nm)/Ni($t_\mathrm{Ni}$) bilayer, $\xi_\mathrm{DL}^E$ increases with $t_\mathrm{Ni}$ and changes sign. This behavior can be interpreted as arising from a competition between spin and orbital torques. 
While the spin torque due to the direct SHE with the negative sign provides the dominant contribution to $\xi_\mathrm{DL}^E$ in the W(3~nm)/Ni(5~nm) bilayer, this contribution remains nearly independent of $t_\mathrm{Ni}$. In contrast, the orbital torque with the positive sign due to the direct OHE increases with $t_\mathrm{Ni}$, leading to the sign reversal of $\xi_\mathrm{DL}^E$. 
A similar competition between spin and orbital torques has been demonstrated in a W/CoFeB structure, where the dominant mechanism depends on the thicknesses of both the W and CoFeB layers~\cite{chiba2024comparative}. The critical thicknesses for the crossover from spin-dominated to orbital-dominated torque are smaller in the W/Ni structure than in the W/CoFeB structure. This difference can be attributed to the stronger orbital response of Ni compared to CoFeB~\cite{Ta-orbital}, which facilitates more efficient orbital torque generation and leads to a lower threshold thickness for the onset of orbital-dominated behavior. 
These results support the interpretation that the observed sign change of $\xi_\mathrm{DL}^E$ arises from the competition between the negative spin torque and the positive orbital torque.

The W layer in the W/Ni bilayer predominantly consists of $\alpha$-W, which is confirmed by the dependence of the W-layer resistivity $\rho_{\rm W}$ on $t_{\rm W}$ (Appendix~B)~\cite{supplementary}. These results indicate that the sign of the bulk direct orbital Hall conductivity $\sigma_\mathrm{OH}^\mathrm{dir,bulk}$ is positive in $\alpha$-W, consistent with theoretical predictions~\cite{PhysRevLett.102.016601,PhysRevMaterials.6.095001,go2024local}.

\begin{figure}[tb]
\includegraphics[scale=1]{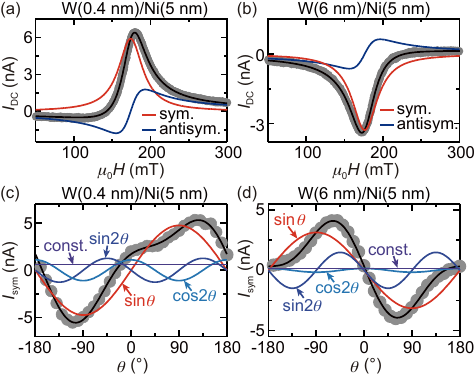}
\caption{Charge-current signals induced by spin/orbital pumping. 
$H$ dependence of $I_{\rm DC}$ for the (a) W(0.4~nm)/Ni(5~nm) and (b) W(6~nm)/Ni(5~nm) bilayers at $\theta=90^\circ$, with $f=10$~GHz and $P_{\rm in}=200$~mW. The solid circles represent the experimental data, and the black solid curves show the fitting results, which are composed of the sum of symmetric (red curve) and antisymmetric (blue curve) components.
In-plane magnetic field angle $\theta$ dependence of $I_{\rm{sym}}$ for the (c) W(0.4~nm)/Ni(5~nm) and (d) W(6~nm)/Ni(5~nm) bilayers. The solid circles represent the experimental data, and the black solid curves show the fitting results using Eq.~(\ref{angle dependence}).
}
\label{fig_pump} 
\end{figure}

In the following, we investigate the orbital-to-charge conversion in W using orbital pumping~\cite{hayashi2024observation,arXiv:2309.14817,han2023theory,PhysRevApplied.19.014069,el2023observation}. The experimental setup and measurement procedures are described in Appendix~C. In the W/Ni bilayer, microwave-driven FMR induces orbital and spin pumping owing to the strong spin-orbit correlation of Ni~\cite{hayashi2024observation,arXiv:2309.14817,han2023theory,PhysRevApplied.19.014069,el2023observation}, injecting orbital and spin currents into the W layer that are converted into charge currents via the inverse OHE and SHE, respectively.
In Figs.~\ref{fig_pump}(a) and \ref{fig_pump}(b), we show the dependence of the charge current $I_{\rm DC}=V_{\rm DC}/R$ on $H$ for the W(0.4~nm)/Ni(5~nm) and W(6~nm)/Ni(5~nm) bilayers, respectively, where $V_{\rm DC}$ is the measured direct-current (DC) voltage, and $R$ denotes the device resistance. By fitting the measured signal using $I_{\rm{DC}}=I_{\rm{sym}}{(\mu_0 W)^2}/[{(\mu_0H-\mu_0H_{\rm{res}})^2+(\mu_0 W)^2}]+I_{\rm{asym}}{(\mu_0 W)(\mu_0H-\mu_0H_{\rm{res}})}/[{(\mu_0 H-\mu_0 H_{\rm{res}})^2+(\mu_0 W)^2}]$, we extract the symmetric component $I_{\rm{sym}}$, where $H_{\rm{res}}$ is the FMR field and $W$ is the FMR linewidth. Here, spin and orbital pumping generate $I_{\rm{sym}}$, while spin rectification effects, including the anisotropic magnetoresistance (AMR), the planar Hall effect (PHE), and the anomalous Hall effect (AHE) in the Ni layer, can contribute to both the symmetric $I_{\rm{sym}}$ and antisymmetric $I_{\rm{asym}}$ components.

To extract the pumping signal $I_\mathrm{pump}$, we measured $I_\mathrm{DC}$ by varying the in-plane magnetic field angle $\theta$, as shown in Figs.~\ref{fig_pump}(c) and \ref{fig_pump}(d). The results are consistent with~\cite{HARDER20161,supplementary}
\begin{equation}
I_{\mathrm{sym}}=I_{\mathrm{pump}}\sin\theta+I_{\mathrm{AMR}}\sin2\theta+I_{\mathrm{PHE}}\cos2\theta +I_{\mathrm{AHE}},
\label{angle dependence}	
\end{equation}
where $I_\mathrm{AMR}$, $I_\mathrm{PHE}$, and $I_\mathrm{AHE}$ are the spin rectification signals due to the AMR, PHE, and AHE, respectively. Figures~\ref{fig_pump}(c) and \ref{fig_pump}(d) show that sizable $I_\mathrm{pump}$ signals are generated by spin and/or orbital pumping in the W/Ni bilayers. 
Using the extracted value of $I_\mathrm{pump}$, we determine the pumping-induced charge-current generation efficiency $\zeta$, defined as~\cite{supplementary}
\begin{equation}
	\zeta  =I_{\mathrm{pump}}\frac{ \gamma \mu_0 H_{\mathrm{res}}\left({\mu_0 W}\right)^2}{2\pi ew  f^2 }\left[1 +\left( \frac{2\pi f}{\gamma\mu_0 H_{\mathrm{res}}}\right)^2\right]^2, 
\end{equation}
where $\gamma$ is the gyromagnetic ratio and $w$ is the width of the device. 
In the model of spin pumping and the inverse SHE, $\zeta$ represents an efficiency parameter that characterizes the combined strength of spin pumping and the inverse SHE as $\zeta = g_{\mathrm{eff}}^{\uparrow \downarrow} \theta_{\mathrm{H}}  \lambda_\mathrm{S} \tanh \left[{t_{\mathrm{W}}}/(2 \lambda_\mathrm{S})\right]  (\mu_0 h)^2$, which varies with the W thickness over the spin diffusion length but should be independent of the Ni thickness.
Here, $g_{\mathrm{eff}}^{\uparrow \downarrow} $ is the effective spin-mixing conductance, and $\theta_{\mathrm{H}}$ is the spin Hall angle. $h$ is the amplitude of the microwave magnetic field, which is consistent across all devices used in this study.

\begin{figure}[tb]
\includegraphics[scale=1]{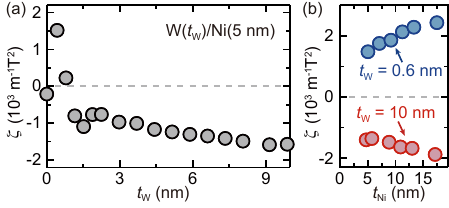}
\caption{The pumping-induced charge-current generation efficiency $\zeta$ in the W/Ni bilayers. (a) Dependence of $\zeta$ on the W thickness $t_\mathrm{W}$ in the W($t_\mathrm{W}$)/Ni(5~nm) bilayer. 
The values of $t_\mathrm{W}$, determined from the sputtering rate, are as follows: 0, 0.42, and 0.79~nm; in 0.36~nm steps from 1.15 to 2.24~nm; in 0.73~nm steps from 2.97 to 8.06~nm; and 9.15 and 9.88~nm.
 (b) Dependence of $\zeta$ on the Ni thickness $t_{\rm Ni}$ in the W(0.6~nm)/Ni($t_\mathrm{Ni}$) (blue) and W(10~nm)/Ni($t_\mathrm{Ni}$) (red) bilayers. The values of $t_\mathrm{Ni}$, determined from the sputtering rate, are as follows: for the W(0.6~nm)/Ni($t_\mathrm{Ni}$), from 5.00 to 13.30~nm in 2.08~nm steps, and 17.45~nm; for the W(10~nm)/Ni($t_\mathrm{Ni}$) structure, 4.68 and 5.72~nm, in 2.08~nm steps from 8.83 to 12.98~nm, and 17.13~nm.}
\label{fig_pumpeff} 
\end{figure}

In Fig.~\ref{fig_pumpeff}(a), we show the dependence of the pumping efficiency $\zeta$ on the W-layer thickness $t_\mathrm{W}$. Notably, $\zeta$ remains negative in the W($t_\mathrm{W}$)/Ni(5~nm) bilayer even for $t_\mathrm{W} > 6$~nm, where the orbital torque efficiency $\xi_\mathrm{DL}^E$ becomes positive (see Fig.~\ref{fig_STFMR}(a)). The negative sign of $\zeta$ corresponds to the negative sign of the charge-current generation mechanism in the W layer. This suggests that $\zeta$ at $t_\mathrm{W} > 6$~nm is dominated by the inverse SHE, which is known to be negative in W. Nevertheless, as shown in Fig.~\ref{fig_pumpeff}(b), $\zeta$ exhibits a strong dependence on the Ni-layer thickness $t_\mathrm{Ni}$ in the W(10~nm)/Ni($t_\mathrm{Ni}$) bilayer. This behavior indicates that $\zeta$ cannot be solely attributed to spin pumping, which should be independent of $t_\mathrm{Ni}$, just as $\xi_\mathrm{DL}^E$ of spin torque---the reciprocal process of spin pumping---is also independent of $t_\mathrm{Ni}$. 
This interpretation is further supported by control experiments on spin-dominated Pt/Ni$_{81}$Fe$_{19}$ bilayers, showing that $\zeta$ is independent of the FM thickness~\cite{supplementary}. We also confirmed that the pumping signal is negligible in Ni single-layer films within this $t_\mathrm{Ni}$ range~\cite{supplementary}.

The observed long-range dependence of $\zeta$ on $t_\mathrm{Ni}$ is characteristic of the reciprocal process of long-range orbital torque, indicating that orbital pumping contributes to the charge current in the W(10~nm)/Ni($t_\mathrm{Ni}$) bilayer.
Notably, the decrease in $\zeta$ with $t_\mathrm{Ni}$ indicates that the orbital-to-charge conversion has a negative sign.
In addition to this negative orbital-to-charge conversion, we also observe a positive orbital-to-charge conversion. As shown in Fig.~\ref{fig_pumpeff}(a), $\zeta$ exhibits a positive peak around $t_\mathrm{W}=0.5$~nm. This positive component increases with $t_\mathrm{Ni}$ as shown in Fig.~\ref{fig_pumpeff}(b), indicating that the signal arises from the orbital-to-charge conversion with the positive sign induced by orbital pumping. 
These results demonstrate that two distinct orbital-to-charge conversion mechanisms with opposite signs coexist in the W($t_\mathrm{W}$)/Ni(5~nm) bilayer: the positive contribution for $t_\mathrm{W} < 1$~nm and the negative contribution for $t_\mathrm{W} > 1$~nm.

The orbital-to-charge conversion with the negative sign for $t_\mathrm{W} > 1$~nm can be attributed to the bulk inverse OHE in the W layer.
The nature of the orbital-to-charge conversion can differ between the bulk and interfaces, accounting for the two orbital-to-charge conversion mechanisms with the opposite signs observed in the W($t_\mathrm{W}$)/Ni(5~nm) bilayer for $t_\mathrm{W} > 1$~nm and $t_\mathrm{W} < 1$~nm. 
Since the interface contribution originates from the anisotropic crystal potential, which differs from the bulk crystal potential~\cite{go2024local}, we assume that the interface contribution is particularly significant at the SiO$_2$/W interface rather than at the metallic W/Ni interface~\cite{supplementary}. This interface effect should be especially strong at small $t_\mathrm{W}$, suggesting that the charge-current generation with the positive sign at $t_\mathrm{W}<1$~nm can be attributed to the orbital-to-charge conversion at the SiO$_2$/W interface. As $t_\mathrm{W}$ increases, this surface contribution is suppressed, while bulk effects, including the bulk inverse SHE and OHE, become more pronounced. This indicates that the bulk inverse OHE dominates the observed orbital-to-charge conversion with the negative sign at larger $t_\mathrm{W}$.

The above pumping results indicate that the bulk inverse OHE has the negative sign ($\sigma_{\mathrm{OH}}^{\mathrm{inv,bulk}} < 0$), which is opposite to the positive sign of the bulk direct OHE ($\sigma_{\mathrm{OH}}^{\mathrm{dir,bulk}} > 0$), as evidenced by the torque measurements. 
These opposite signs clearly demonstrate a violation of local reciprocity, as expressed by Eq.~(\ref{local}), in the charge-orbital interconversion in $\alpha$-W. The signs of the bulk direct and inverse OHEs, obtained directly from our thickness-dependent torque and pumping measurements, are consistent with the theoretical prediction summarized in Table~I in Appendix~A.
Table~I also suggests that the surface direct OHE with a negative sign may contribute to the negative torque for $t_\mathrm{W} < 4$~nm, although its sign cannot be determined solely from the experimental data~\cite{supplementary}. Importantly, even though interfacial contributions cannot be fully quantified, the bulk contributions are reliably identified from the thickness dependence of the torque and pumping signals. Therefore, the key conclusion of this work---the direct and inverse OHEs in the bulk exhibit the opposite signs---remains robust. 
The observed local nonreciprocity in the OHEs contrasts sharply with the SHEs, where the direct and inverse SHEs remain largely reciprocal even at a local level in $\alpha$-W~\cite{go2024local}. This contrast highlights the distinct nature of charge-orbital interconversion compared with charge-spin interconversion.

In summary, we have experimentally demonstrated the violation of local reciprocity in the charge-orbital interconversion in the W/Ni bilayers. By measuring the orbital torque and orbital pumping, we found that the sign of the bulk inverse OHE is opposite to that of the bulk direct OHE. This result confirms that the charge-orbital interconversion is locally nonreciprocal, $\sigma_{\mathrm{OH}}^{\mathrm{dir}}(z) \neq \sigma_{\mathrm{OH}}^{\mathrm{inv}}(z)$, in sharp contrast to the global reciprocity expressed in Eq.~(\ref{global}), which is governed by Onsager's reciprocal relations. 
The local violation of reciprocity between the direct and inverse OHEs is also expected in other heavy and light metals~\cite{supplementary}, suggesting that local nonreciprocity could serve as a powerful experimental signature of the OHE across a wide range of materials. Because local nonreciprocity originates from the nonconservation of OAM, extending such studies to systems where the loss of OAM can be controlled through interactions with magnons or phonons could open new directions for manipulating charge-orbital coupled transport. In spintronics, while the direct and inverse SHEs have generally been regarded as reciprocal processes, recent studies have revealed that nonreciprocal charge-spin conversion can emerge in Rashba systems or through spin-vorticity coupling~\cite{PhysRevLett.122.217701,PhysRevB.96.140409}. These findings highlight that investigating reciprocity in charge-spin and charge-orbital interconversions is essential for achieving a unified understanding of angular-momentum transport in solids.

\begin{acknowledgments}
We thank J. Yoon and T. Gao for valuable discussions, and K. Yamanoi and Y. Nozaki for their assistance in sample preparation.
This work was supported by JSPS KAKENHI (Grant Numbers: 22H04964, 23K19037, 25K21707), Spintronics Research Network of Japan (Spin-RNJ), and MEXT Initiative to Establish Next-generation Novel Integrated Circuits Centers (X-NICS) (Grant Number: JPJ011438). 
Y.M. and D.G. acknowledge support from the EIC Pathfinder OPEN grant 101129641 ``OBELIX".
\end{acknowledgments}

\clearpage

\onecolumngrid
\subsection{End Matter}
\twocolumngrid

\textit{Appendix A: Local nonreciprocity}---Table~I summarizes the theoretically predicted bulk and surface contributions to the direct and inverse OHEs in $\alpha$-W~\cite{go2024local}. The results show that the local responses exhibit opposite signs between the direct and inverse OHEs, indicating a violation of local reciprocity, $\sigma_{\mathrm{OH}}^{\mathrm{dir}}(z) \neq \sigma_{\mathrm{OH}}^{\mathrm{inv}}(z)$. 
This prediction indicates that while the direct and inverse OHEs are globally reciprocal, their reciprocity can be locally violated. This is in stark contrast to the SHEs, which has been predicted to be reciprocal even locally~\cite{go2024local}. Such a distinction is of fundamental importance, as it provides a key experimental signature for differentiating between spin and orbital currents.

\begin{center}
  \begin{table}[tb]
 \caption{Decomposition of the bulk and surface contributions to the direct and inverse OHEs in $\alpha$-W~\cite{go2024local}. The bulk and surface contributions to the direct(inverse) orbital Hall conductivity are represented as $\sigma_{\mathrm{OH}}^{\mathrm{dir(inv), bulk}} $ and $\sigma_{\mathrm{OH}}^{\mathrm{dir(inv), surface}} $, respectively, along with the spin Hall conductivity components represented as $\sigma_{\mathrm{SH}}^{\mathrm{dir(inv), bulk}} $ and $\sigma_{\mathrm{SH}}^{\mathrm{dir(inv), surface}} $. 
 }
   \label{table1}
   \centering
     \renewcommand{\arraystretch}{1.3}                  
    \begin{tabular}{cccc}
     \hline \hline
  Orbital/Spin&     Conversion  &  Bulk    &   Surface \\
     \hline
 OHE & Direct  & $\sigma_{\mathrm{OH}}^{\mathrm{dir, bulk}} >0$  & $\sigma_{\mathrm{OH}}^{\mathrm{dir, surface}} <0$  \\
 & Inverse   &  $\sigma_{\mathrm{OH}}^{\mathrm{inv, bulk}} <0$ &  $\sigma_{\mathrm{OH}}^{\mathrm{inv, surface}} >0$ \\
  \hline
 SHE & Direct & $\sigma_{\mathrm{SH}}^{\mathrm{dir, bulk}} <0$  & $\sigma_{\mathrm{SH}}^{\mathrm{dir, surface}} \simeq 0$  \\
 & Inverse &  $\sigma_{\mathrm{SH}}^{\mathrm{inv, bulk}} <0$ &  $\sigma_{\mathrm{SH}}^{\mathrm{inv, surface}} \simeq 0$ \\
           \hline 
          \hline
    \end{tabular}
  \end{table}
\end{center}

\begin{figure}[tb]
\includegraphics[scale=1]{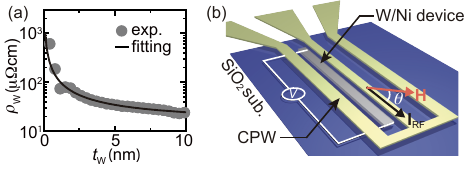}
\caption{
(a) Dependence of the W-layer resistivity $\rho_\mathrm{W}$ on $t_{\rm W}$ in the Ni(5~nm)/W($t_\mathrm{W}$) bilayer. The solid circles are the experimental data, and the solid curve is the fitting result.
(b) Schematic illustration of the experimental setup used to measure the spin/orbital pumping. ${\bf H}$ and ${\bf I}_\mathrm{RF}$ denote the external magnetic field and the RF current, respectively. $\theta$ is defined as the in-plane angle between ${\bf H}$ and ${\bf I}_\mathrm{RF}$. 
}
\label{fig_appendix} 
\end{figure}

\textit{Appendix B: Resistivity of W}---The W layer in the W/Ni bilayer predominantly consists of $\alpha$-W, which is confirmed by the dependence of the W-layer resistivity $\rho_{\rm W}$ on $t_{\rm W}$, shown in Fig.~\ref{fig_appendix}(a)~\cite{supplementary}. Figure~\ref{fig_appendix}(a) shows that $\rho_{\rm W}$ follows $ \rho_{\rm W}(t_{\rm W}) = a t_{\rm W}^{-1} + \rho_{\rm W}^{\rm bulk}$, where $a t_{\rm W}^{-1}$ accounts for the resistivity contribution from surface scattering, and $\rho_{\rm W}^{\rm bulk}$ is the bulk-limit resistivity. The extracted value of $\rho_{\rm W}^{\rm bulk} = 17~\mu\Omega \text{cm}$ is consistent with the resistivity of low-resistivity $\alpha$-W~\cite{hayashi2022observation}.

\vspace{8pt}

\textit{Appendix C: Pumping measurements}---For the pumping measurements, the W/Ni film was patterned into rectangular strips with a width of $8$~$\mu$m and a length of $180$~$\mu$m using photolithography and Ar-ion milling. The device was positioned between the ground and signal lines of a Au(400 nm)/Ti(3 nm) coplanar waveguide (CPW), with a gap of 8~$\mu$m separating the CPW from the device (see Fig.~\ref{fig_appendix}(b)). During the measurements, a radio frequency (RF) current $I_\mathrm{RF}$ with a frequency $f$ and a power $P_\mathrm{in}$ was applied to the CPW, along with an in-plane external magnetic field $H$ at an angle $\theta$ relative to the longitudinal direction of the device. The RF current generates an out-of-plane microwave magnetic field, driving FMR in the W/Ni bilayer. We measured the direct-current (DC) voltage $V_{\rm DC}$ induced by the FMR across electrodes attached to the W/Ni bilayer.


\begin{thebibliography}{10}
\expandafter\ifx\csname url\endcsname\relax
  \def\url#1{\texttt{#1}}\fi
\expandafter\ifx\csname urlprefix\endcsname\relax\def\urlprefix{URL }\fi
\providecommand{\bibinfo}[2]{#2}
\providecommand{\eprint}[2][]{\url{#2}}

\bibitem{RevModPhys.87.1213}
\bibinfo{author}{Sinova, J.}, \bibinfo{author}{Valenzuela, S.~O.},
  \bibinfo{author}{Wunderlich, J.}, \bibinfo{author}{Back, C.~H.} \&
  \bibinfo{author}{Jungwirth, T.}
\newblock \bibinfo{title}{{Spin Hall effects}}.
\newblock \emph{\bibinfo{journal}{Rev. Mod. Phys.}}
  \textbf{\bibinfo{volume}{87}}, \bibinfo{pages}{1213--1260}
  (\bibinfo{year}{2015}).

\bibitem{AndoPRL}
\bibinfo{author}{Ando, K.} \emph{et~al.}
\newblock \bibinfo{title}{{Electric manipulation of spin relaxation using the
  spin Hall effect}}.
\newblock \emph{\bibinfo{journal}{Phys. Rev. Lett.}}
  \textbf{\bibinfo{volume}{101}}, \bibinfo{pages}{036601}
  (\bibinfo{year}{2008}).

\bibitem{liu2012spin}
\bibinfo{author}{Liu, L.} \emph{et~al.}
\newblock \bibinfo{title}{{Spin-torque switching with the giant spin Hall
  effect of tantalum}}.
\newblock \emph{\bibinfo{journal}{Science}} \textbf{\bibinfo{volume}{336}},
  \bibinfo{pages}{555--558} (\bibinfo{year}{2012}).

\bibitem{RevModPhys.91.035004}
\bibinfo{author}{Manchon, A.} \emph{et~al.}
\newblock \bibinfo{title}{Current-induced spin-orbit torques in ferromagnetic
  and antiferromagnetic systems}.
\newblock \emph{\bibinfo{journal}{Rev. Mod. Phys.}}
  \textbf{\bibinfo{volume}{91}}, \bibinfo{pages}{035004}
  (\bibinfo{year}{2019}).

\bibitem{saitoh2006conversion}
\bibinfo{author}{Saitoh, E.}, \bibinfo{author}{Ueda, M.},
  \bibinfo{author}{Miyajima, H.} \& \bibinfo{author}{Tatara, G.}
\newblock \bibinfo{title}{Conversion of spin current into charge current at
  room temperature: Inverse spin-hall effect}.
\newblock \emph{\bibinfo{journal}{Appl. Phys. Lett.}}
  \textbf{\bibinfo{volume}{88}}, \bibinfo{pages}{182509}
  (\bibinfo{year}{2006}).

\bibitem{valenzuela2006direct}
\bibinfo{author}{Valenzuela, S.~O.} \& \bibinfo{author}{Tinkham, M.}
\newblock \bibinfo{title}{Direct electronic measurement of the spin hall
  effect}.
\newblock \emph{\bibinfo{journal}{Nature}} \textbf{\bibinfo{volume}{442}},
  \bibinfo{pages}{176--179} (\bibinfo{year}{2006}).

\bibitem{KimuraPRL}
\bibinfo{author}{Kimura, T.}, \bibinfo{author}{Otani, Y.},
  \bibinfo{author}{Sato, T.}, \bibinfo{author}{Takahashi, S.} \&
  \bibinfo{author}{Maekawa, S.}
\newblock \bibinfo{title}{{Room-temperature reversible spin Hall effect}}.
\newblock \emph{\bibinfo{journal}{Phys. Rev. Lett.}}
  \textbf{\bibinfo{volume}{98}}, \bibinfo{pages}{156601}
  (\bibinfo{year}{2007}).

\bibitem{PhysRevLett.95.066601}
\bibinfo{author}{Bernevig, B.~A.}, \bibinfo{author}{Hughes, T.~L.} \&
  \bibinfo{author}{Zhang, S.-C.}
\newblock \bibinfo{title}{Orbitronics: The intrinsic orbital current in
  $p$-doped silicon}.
\newblock \emph{\bibinfo{journal}{Phys. Rev. Lett.}}
  \textbf{\bibinfo{volume}{95}}, \bibinfo{pages}{066601}
  (\bibinfo{year}{2005}).

\bibitem{PhysRevLett.102.016601}
\bibinfo{author}{Kontani, H.}, \bibinfo{author}{Tanaka, T.},
  \bibinfo{author}{Hirashima, D.~S.}, \bibinfo{author}{Yamada, K.} \&
  \bibinfo{author}{Inoue, J.}
\newblock \bibinfo{title}{{Giant orbital Hall effect in transition metals:
  Origin of large spin and anomalous Hall effects}}.
\newblock \emph{\bibinfo{journal}{Phys. Rev. Lett.}}
  \textbf{\bibinfo{volume}{102}}, \bibinfo{pages}{016601}
  (\bibinfo{year}{2009}).

\bibitem{PhysRevLett.121.086602}
\bibinfo{author}{Go, D.}, \bibinfo{author}{Jo, D.}, \bibinfo{author}{Kim, C.}
  \& \bibinfo{author}{Lee, H.-W.}
\newblock \bibinfo{title}{{Intrinsic spin and orbital Hall effects from orbital
  texture}}.
\newblock \emph{\bibinfo{journal}{Phys. Rev. Lett.}}
  \textbf{\bibinfo{volume}{121}}, \bibinfo{pages}{086602}
  (\bibinfo{year}{2018}).

\bibitem{PhysRevB.102.035409}
\bibinfo{author}{Bhowal, S.} \& \bibinfo{author}{Satpathy, S.}
\newblock \bibinfo{title}{{Intrinsic orbital and spin Hall effects in monolayer
  transition metal dichalcogenides}}.
\newblock \emph{\bibinfo{journal}{Phys. Rev. B}}
  \textbf{\bibinfo{volume}{102}}, \bibinfo{pages}{035409}
  (\bibinfo{year}{2020}).

\bibitem{PhysRevLett.126.056601}
\bibinfo{author}{Cysne, T.~P.} \emph{et~al.}
\newblock \bibinfo{title}{{Disentangling orbital and valley Hall effects in
  bilayers of transition metal dichalcogenides}}.
\newblock \emph{\bibinfo{journal}{Phys. Rev. Lett.}}
  \textbf{\bibinfo{volume}{126}}, \bibinfo{pages}{056601}
  (\bibinfo{year}{2021}).

\bibitem{go2021orbitronics}
\bibinfo{author}{Go, D.}, \bibinfo{author}{Jo, D.}, \bibinfo{author}{Lee,
  H.-W.}, \bibinfo{author}{Kl{\"a}ui, M.} \& \bibinfo{author}{Mokrousov, Y.}
\newblock \bibinfo{title}{{Orbitronics: Orbital currents in solids}}.
\newblock \emph{\bibinfo{journal}{EPL}} \textbf{\bibinfo{volume}{135}},
  \bibinfo{pages}{37001} (\bibinfo{year}{2021}).

\bibitem{KIM2022169974}
\bibinfo{author}{Kim, J.} \& \bibinfo{author}{Otani, Y.}
\newblock \bibinfo{title}{Orbital angular momentum for spintronics}.
\newblock \emph{\bibinfo{journal}{J. Magn. Magn. Mater.}}
  \textbf{\bibinfo{volume}{563}}, \bibinfo{pages}{169974}
  (\bibinfo{year}{2022}).

\bibitem{2025Andoreview}
\bibinfo{author}{Ando, K.}
\newblock \bibinfo{title}{Orbitronics: Harnessing orbital currents in
  solid-state devices}.
\newblock \emph{\bibinfo{journal}{J. Phys. Soc. Jpn.}}
  \textbf{\bibinfo{volume}{94}}, \bibinfo{pages}{092001}
  (\bibinfo{year}{2025}).

\bibitem{PhysRevLett.132.226704}
\bibinfo{author}{Mendoza-Rodarte, J.~A.}, \bibinfo{author}{Cosset-Ch\'eneau,
  M.}, \bibinfo{author}{van Wees, B.~J.} \& \bibinfo{author}{Guimar\~aes, M.
  H.~D.}
\newblock \bibinfo{title}{{Efficient magnon injection and detection via the
  orbital Rashba-Edelstein effect}}.
\newblock \emph{\bibinfo{journal}{Phys. Rev. Lett.}}
  \textbf{\bibinfo{volume}{132}}, \bibinfo{pages}{226704}
  (\bibinfo{year}{2024}).

\bibitem{4c06056}
\bibinfo{author}{Ledesma-Martin, J.} \emph{et~al.}
\newblock \bibinfo{title}{Nonreciprocity in magnon mediated charge-spin-orbital
  current interconversion}.
\newblock \emph{\bibinfo{journal}{Nano Lett.}} \textbf{\bibinfo{volume}{25}},
  \bibinfo{pages}{3247--3252} (\bibinfo{year}{2025}).

\bibitem{gao2025nonlocalelectricaldetectionreciprocal}
\bibinfo{author}{Gao, W.} \emph{et~al.}
\newblock \bibinfo{title}{{Nonlocal electrical detection of reciprocal orbital
  Edelstein effect}}.
\newblock \emph{\bibinfo{journal}{Nat. Commun.}} \textbf{\bibinfo{volume}{16}},
  \bibinfo{pages}{6380} (\bibinfo{year}{2025}).

\bibitem{supplementary}
\bibinfo{note}{See Supplemental Material for the detailed descriptions of
  Global and local reciprocity, Material choice for investigating
  nonreciprocity, Structural characterization of W, Spin-torque ferromagnetic
  resonance, Definition of pumping-induced charge-current generation
  efficiency, Separating spin rectification contributions, Charge-current
  signals in Ni single-layer films, Torque and pumping measurements in
  reference systems, Summary of experimental results, Interfacial
  contributions, Sign reversal of torque and pumping signals, Bulk and surface
  orbital Hall effects in polycrystalline films, which includes Refs. [20-30].}

\bibitem{lee2024composition}
\bibinfo{author}{Lee, H.} \& \bibinfo{author}{Lee, H.-W.}
\newblock \bibinfo{title}{{Composition dependence of the orbital torque in
  Co$_x$Fe$_{1- x}$ and Ni$_x$Fe$_{1- x}$ alloys: Spin-orbit correlation
  analysis}}.
\newblock \emph{\bibinfo{journal}{Curr. Appl. Phys.}}
  \textbf{\bibinfo{volume}{67}}, \bibinfo{pages}{60--68}
  (\bibinfo{year}{2024}).

\bibitem{10.1116/1.5003628}
\bibinfo{author}{Barmak, K.} \& \bibinfo{author}{Liu, J.}
\newblock \bibinfo{title}{Impact of deposition rate, underlayers, and
  substrates on $\beta$-tungsten formation in sputter deposited films}.
\newblock \emph{\bibinfo{journal}{J. Vac. Sci. Technol. A}}
  \textbf{\bibinfo{volume}{35}}, \bibinfo{pages}{061516}
  (\bibinfo{year}{2017}).

\bibitem{vullers2015alpha}
\bibinfo{author}{V{\"u}llers, F.} \& \bibinfo{author}{Spolenak, R.}
\newblock \bibinfo{title}{{Alpha-vs. beta-W nanocrystalline thin films: A
  comprehensive study of sputter parameters and resulting materials'
  properties}}.
\newblock \emph{\bibinfo{journal}{Thin Solid Films}}
  \textbf{\bibinfo{volume}{577}}, \bibinfo{pages}{26--34}
  (\bibinfo{year}{2015}).

\bibitem{liao2019current}
\bibinfo{author}{Liao, W.-B.}, \bibinfo{author}{Chen, T.-Y.},
  \bibinfo{author}{Ferrante, Y.}, \bibinfo{author}{Parkin, S.~S.} \&
  \bibinfo{author}{Pai, C.-F.}
\newblock \bibinfo{title}{{Current-induced magnetization switching by the high
  spin Hall conductivity $\alpha$-W}}.
\newblock \emph{\bibinfo{journal}{Phys. Status Solidi RRL}}
  \textbf{\bibinfo{volume}{13}}, \bibinfo{pages}{1900408}
  (\bibinfo{year}{2019}).

\bibitem{kawada2021acoustic}
\bibinfo{author}{Kawada, T.}, \bibinfo{author}{Kawaguchi, M.},
  \bibinfo{author}{Funato, T.}, \bibinfo{author}{Kohno, H.} \&
  \bibinfo{author}{Hayashi, M.}
\newblock \bibinfo{title}{{Acoustic spin Hall effect in strong spin-orbit
  metals}}.
\newblock \emph{\bibinfo{journal}{Sci. Adv.}} \textbf{\bibinfo{volume}{7}},
  \bibinfo{pages}{eabd9697} (\bibinfo{year}{2021}).

\bibitem{liu2015correlation}
\bibinfo{author}{Liu, J.}, \bibinfo{author}{Ohkubo, T.},
  \bibinfo{author}{Mitani, S.}, \bibinfo{author}{Hono, K.} \&
  \bibinfo{author}{Hayashi, M.}
\newblock \bibinfo{title}{{Correlation between the spin Hall angle and the
  structural phases of early 5$d$ transition metals}}.
\newblock \emph{\bibinfo{journal}{Appl. Phys. Lett.}}
  \textbf{\bibinfo{volume}{107}} (\bibinfo{year}{2015}).

\bibitem{fang2011spin}
\bibinfo{author}{Fang, D.} \emph{et~al.}
\newblock \bibinfo{title}{Spin-orbit-driven ferromagnetic resonance}.
\newblock \emph{\bibinfo{journal}{Nat. Nanotechnol.}}
  \textbf{\bibinfo{volume}{6}}, \bibinfo{pages}{413--417}
  (\bibinfo{year}{2011}).

\bibitem{PhysRevB.92.214406}
\bibinfo{author}{Tshitoyan, V.} \emph{et~al.}
\newblock \bibinfo{title}{{Electrical manipulation of ferromagnetic NiFe by
  antiferromagnetic IrMn}}.
\newblock \emph{\bibinfo{journal}{Phys. Rev. B}} \textbf{\bibinfo{volume}{92}},
  \bibinfo{pages}{214406} (\bibinfo{year}{2015}).

\bibitem{moriya2024nano}
\bibinfo{author}{Moriya, H.} \emph{et~al.}
\newblock \bibinfo{title}{{Observation of long-range current-induced torque in
  Ni/Pt bilayers}}.
\newblock \emph{\bibinfo{journal}{Nano Lett.}} \textbf{\bibinfo{volume}{24}},
  \bibinfo{pages}{6459--6464} (\bibinfo{year}{2024}).

\bibitem{PhysRevB.90.224427}
\bibinfo{author}{Avci, C.~O.} \emph{et~al.}
\newblock \bibinfo{title}{Interplay of spin-orbit torque and thermoelectric
  effects in ferromagnet/normal-metal bilayers}.
\newblock \emph{\bibinfo{journal}{Phys. Rev. B}} \textbf{\bibinfo{volume}{90}},
  \bibinfo{pages}{224427} (\bibinfo{year}{2014}).

\bibitem{adma.202301608}
\bibinfo{author}{Patton, M.} \emph{et~al.}
\newblock \bibinfo{title}{{Symmetry control of unconventional spin-orbit
  torques in IrO$_2$}}.
\newblock \emph{\bibinfo{journal}{Adv. Mater.}} \textbf{\bibinfo{volume}{35}},
  \bibinfo{pages}{2301608} (\bibinfo{year}{2023}).

\bibitem{go2024local}
\bibinfo{author}{Go, D.} \emph{et~al.}
\newblock \bibinfo{title}{Local and global reciprocity in
  orbital-charge-coupled transport}.
\newblock \emph{\bibinfo{journal}{arXiv:2407.00517}}  (\bibinfo{year}{2024}).

\bibitem{Cr-orbital}
\bibinfo{author}{Lee, S.} \emph{et~al.}
\newblock \bibinfo{title}{{Efficient conversion of orbital Hall current to spin
  current for spin-orbit torque switching}}.
\newblock \emph{\bibinfo{journal}{Commun. Phys.}} \textbf{\bibinfo{volume}{4}},
  \bibinfo{pages}{234} (\bibinfo{year}{2021}).

\bibitem{Ta-orbital}
\bibinfo{author}{Lee, D.} \emph{et~al.}
\newblock \bibinfo{title}{Orbital torque in magnetic bilayers}.
\newblock \emph{\bibinfo{journal}{Nat. Commun.}} \textbf{\bibinfo{volume}{12}},
  \bibinfo{pages}{6710} (\bibinfo{year}{2021}).

\bibitem{PhysRevResearch.4.033037}
\bibinfo{author}{Sala, G.} \& \bibinfo{author}{Gambardella, P.}
\newblock \bibinfo{title}{{Giant orbital Hall effect and orbital-to-spin
  conversion in $3d$, $5d$, and $4f$ metallic heterostructures}}.
\newblock \emph{\bibinfo{journal}{Phys. Rev. Research}}
  \textbf{\bibinfo{volume}{4}}, \bibinfo{pages}{033037} (\bibinfo{year}{2022}).

\bibitem{hayashi2022observation}
\bibinfo{author}{Hayashi, H.} \emph{et~al.}
\newblock \bibinfo{title}{Observation of long-range orbital transport and giant
  orbital torque}.
\newblock \emph{\bibinfo{journal}{Commun. Phys.}} \textbf{\bibinfo{volume}{6}},
  \bibinfo{pages}{32} (\bibinfo{year}{2023}).

\bibitem{choi2021observation}
\bibinfo{author}{Choi, Y.-G.} \emph{et~al.}
\newblock \bibinfo{title}{{Observation of the orbital Hall effect in a light
  metal Ti}}.
\newblock \emph{\bibinfo{journal}{Nature}} \textbf{\bibinfo{volume}{619}},
  \bibinfo{pages}{52--56} (\bibinfo{year}{2023}).

\bibitem{hayashi2024observation}
\bibinfo{author}{Hayashi, H.}, \bibinfo{author}{Go, D.}, \bibinfo{author}{Haku,
  S.}, \bibinfo{author}{Mokrousov, Y.} \& \bibinfo{author}{Ando, K.}
\newblock \bibinfo{title}{Observation of orbital pumping}.
\newblock \emph{\bibinfo{journal}{Nat. Electron.}}
  \textbf{\bibinfo{volume}{7}}, \bibinfo{pages}{646--652}
  (\bibinfo{year}{2024}).

\bibitem{hayashi2023orbital}
\bibinfo{author}{Hayashi, H.} \& \bibinfo{author}{Ando, K.}
\newblock \bibinfo{title}{{Orbital Hall magnetoresistance in Ni/Ti bilayers}}.
\newblock \emph{\bibinfo{journal}{Appl. Phys. Lett.}}
  \textbf{\bibinfo{volume}{123}} (\bibinfo{year}{2023}).

\bibitem{PhysRevB.107.134423}
\bibinfo{author}{Bose, A.} \emph{et~al.}
\newblock \bibinfo{title}{{Detection of long-range orbital-Hall torques}}.
\newblock \emph{\bibinfo{journal}{Phys. Rev. B}}
  \textbf{\bibinfo{volume}{107}}, \bibinfo{pages}{134423}
  (\bibinfo{year}{2023}).

\bibitem{xu2024orbitronics}
\bibinfo{author}{Xu, Y.} \emph{et~al.}
\newblock \bibinfo{title}{{Orbitronics: light-induced orbital currents in Ni
  studied by terahertz emission experiments}}.
\newblock \emph{\bibinfo{journal}{Nat. Commun.}} \textbf{\bibinfo{volume}{15}},
  \bibinfo{pages}{2043} (\bibinfo{year}{2024}).

\bibitem{seifert2023time}
\bibinfo{author}{Seifert, T.~S.} \emph{et~al.}
\newblock \bibinfo{title}{Time-domain observation of ballistic
  orbital-angular-momentum currents with giant relaxation length in tungsten}.
\newblock \emph{\bibinfo{journal}{Nat. Nanotechnol.}}
  \textbf{\bibinfo{volume}{18}}, \bibinfo{pages}{1132--1138}
  (\bibinfo{year}{2023}).

\bibitem{gao2024control}
\bibinfo{author}{Gao, T.} \emph{et~al.}
\newblock \bibinfo{title}{Control of dynamic orbital response in ferromagnets
  via crystal symmetry}.
\newblock \emph{\bibinfo{journal}{Nat. Phys.}} \textbf{\bibinfo{volume}{20}},
  \bibinfo{pages}{1896--1903} (\bibinfo{year}{2024}).

\bibitem{PhysRevLett.106.036601}
\bibinfo{author}{Liu, L.}, \bibinfo{author}{Moriyama, T.},
  \bibinfo{author}{Ralph, D.~C.} \& \bibinfo{author}{Buhrman, R.~A.}
\newblock \bibinfo{title}{{Spin-torque ferromagnetic resonance induced by the
  spin Hall effect}}.
\newblock \emph{\bibinfo{journal}{Phys. Rev. Lett.}}
  \textbf{\bibinfo{volume}{106}}, \bibinfo{pages}{036601}
  (\bibinfo{year}{2011}).

\bibitem{PhysRevMaterials.6.095001}
\bibinfo{author}{Salemi, L.} \& \bibinfo{author}{Oppeneer, P.~M.}
\newblock \bibinfo{title}{{First-principles theory of intrinsic spin and
  orbital Hall and Nernst effects in metallic monoatomic crystals}}.
\newblock \emph{\bibinfo{journal}{Phys. Rev. Materials}}
  \textbf{\bibinfo{volume}{6}}, \bibinfo{pages}{095001} (\bibinfo{year}{2022}).

\bibitem{PhysRevMaterials.2.014403}
\bibinfo{author}{Wang, T.-C.}, \bibinfo{author}{Chen, T.-Y.},
  \bibinfo{author}{Wu, C.-T.}, \bibinfo{author}{Yen, H.-W.} \&
  \bibinfo{author}{Pai, C.-F.}
\newblock \bibinfo{title}{{Comparative study on spin-orbit torque efficiencies
  from W/ferromagnetic and W/ferrimagnetic heterostructures}}.
\newblock \emph{\bibinfo{journal}{Phys. Rev. Materials}}
  \textbf{\bibinfo{volume}{2}}, \bibinfo{pages}{014403} (\bibinfo{year}{2018}).

\bibitem{PhysRevLett.130.246701}
\bibinfo{author}{Go, D.} \emph{et~al.}
\newblock \bibinfo{title}{Long-range orbital torque by momentum-space
  hotspots}.
\newblock \emph{\bibinfo{journal}{Phys. Rev. Lett.}}
  \textbf{\bibinfo{volume}{130}}, \bibinfo{pages}{246701}
  (\bibinfo{year}{2023}).

\bibitem{PhysRevLett.128.067201}
\bibinfo{author}{Ding, S.} \emph{et~al.}
\newblock \bibinfo{title}{{Observation of the orbital Rashba-Edelstein
  magnetoresistance}}.
\newblock \emph{\bibinfo{journal}{Phys. Rev. Lett.}}
  \textbf{\bibinfo{volume}{128}}, \bibinfo{pages}{067201}
  (\bibinfo{year}{2022}).

\bibitem{PhysRevB.105.104434}
\bibinfo{author}{Liao, L.} \emph{et~al.}
\newblock \bibinfo{title}{{Efficient orbital torque in polycrystalline
  $\mathrm{ferromagnetic}\text{\ensuremath{-}}\mathrm{metal}/\mathrm{Ru}/{\mathrm{Al}}_{2}{\mathrm{O}}_{3}$
  stacks: Theory and experiment}}.
\newblock \emph{\bibinfo{journal}{Phys. Rev. B}}
  \textbf{\bibinfo{volume}{105}}, \bibinfo{pages}{104434}
  (\bibinfo{year}{2022}).

\bibitem{PhysRevResearch.5.023054}
\bibinfo{author}{Fukunaga, R.}, \bibinfo{author}{Haku, S.},
  \bibinfo{author}{Hayashi, H.} \& \bibinfo{author}{Ando, K.}
\newblock \bibinfo{title}{{Orbital torque originating from orbital Hall effect
  in Zr}}.
\newblock \emph{\bibinfo{journal}{Phys. Rev. Res.}}
  \textbf{\bibinfo{volume}{5}}, \bibinfo{pages}{023054} (\bibinfo{year}{2023}).

\bibitem{chiba2024comparative}
\bibinfo{author}{Chiba, S.}, \bibinfo{author}{Marui, Y.},
  \bibinfo{author}{Ohno, H.} \& \bibinfo{author}{Fukami, S.}
\newblock \bibinfo{title}{{Comparative study of current-induced torque in
  Cr/CoFeB/MgO and W/CoFeB/MgO}}.
\newblock \emph{\bibinfo{journal}{Nano Lett.}} \textbf{\bibinfo{volume}{24}},
  \bibinfo{pages}{14028--14033} (\bibinfo{year}{2024}).

\bibitem{arXiv:2309.14817}
\bibinfo{author}{Go, D.} \emph{et~al.}
\newblock \bibinfo{title}{Orbital pumping by magnetization dynamics in
  ferromagnets}.
\newblock \emph{\bibinfo{journal}{arXiv:2309.14817}}  (\bibinfo{year}{2023}).

\bibitem{han2023theory}
\bibinfo{author}{Han, S.} \emph{et~al.}
\newblock \bibinfo{title}{Orbital pumping incorporating both orbital angular
  momentum and position}.
\newblock \emph{\bibinfo{journal}{Phys. Rev. Lett.}}
  \textbf{\bibinfo{volume}{134}}, \bibinfo{pages}{036305}
  (\bibinfo{year}{2025}).

\bibitem{PhysRevApplied.19.014069}
\bibinfo{author}{Santos, E.} \emph{et~al.}
\newblock \bibinfo{title}{Inverse orbital torque via spin-orbital intertwined
  states}.
\newblock \emph{\bibinfo{journal}{Phys. Rev. Appl.}}
  \textbf{\bibinfo{volume}{19}}, \bibinfo{pages}{014069}
  (\bibinfo{year}{2023}).

\bibitem{el2023observation}
\bibinfo{author}{El~Hamdi, A.} \emph{et~al.}
\newblock \bibinfo{title}{{Observation of the orbital inverse Rashba--Edelstein
  effect}}.
\newblock \emph{\bibinfo{journal}{Nat. Phys.}} \textbf{\bibinfo{volume}{19}},
  \bibinfo{pages}{1855--1860} (\bibinfo{year}{2023}).

\bibitem{HARDER20161}
\bibinfo{author}{Harder, M.}, \bibinfo{author}{Gui, Y.} \& \bibinfo{author}{Hu,
  C.-M.}
\newblock \bibinfo{title}{Electrical detection of magnetization dynamics via
  spin rectification effects}.
\newblock \emph{\bibinfo{journal}{Phys. Rep.}} \textbf{\bibinfo{volume}{661}},
  \bibinfo{pages}{1--59} (\bibinfo{year}{2016}).

\bibitem{PhysRevLett.122.217701}
\bibinfo{author}{Okano, G.}, \bibinfo{author}{Matsuo, M.},
  \bibinfo{author}{Ohnuma, Y.}, \bibinfo{author}{Maekawa, S.} \&
  \bibinfo{author}{Nozaki, Y.}
\newblock \bibinfo{title}{Nonreciprocal spin current generation in
  surface-oxidized copper films}.
\newblock \emph{\bibinfo{journal}{Phys. Rev. Lett.}}
  \textbf{\bibinfo{volume}{122}}, \bibinfo{pages}{217701}
  (\bibinfo{year}{2019}).

\bibitem{PhysRevB.96.140409}
\bibinfo{author}{Kim, J.} \emph{et~al.}
\newblock \bibinfo{title}{Evaluation of bulk-interface contributions to
  edelstein magnetoresistance at metal/oxide interfaces}.
\newblock \emph{\bibinfo{journal}{Phys. Rev. B}} \textbf{\bibinfo{volume}{96}},
  \bibinfo{pages}{140409} (\bibinfo{year}{2017}).

\end{thebibliography}
\end{document}